\documentclass[preprint,12pt]{elsarticle}

\usepackage{amssymb}
\usepackage{amsmath}

\journal{Nuclear Physics A}

\begin{document}

\begin{frontmatter}



\title{A Community-Enhanced Graph Representation Model for Link Prediction}


\author[1]{Lei Wang}\ead{lei_wang@seu.edu.cn}
\author[1]{Darong Lai\corref{cor1}}\ead{daronglai@seu.edu.cn}
\cortext[cor1]{Corresponding author.}

\affiliation[1]{organization={School of Computer Science and Engineering },
                addressline={Southeast University},
                city={Nanjing},
                postcode={211189},
                country={China}}

\begin{abstract}
Although Graph Neural Networks (GNNs) have become the dominant approach for graph representation learning, their performance on link prediction tasks does not always surpass that of traditional heuristic methods such as Common Neighbors and Jaccard Coefficient. This is mainly because existing GNNs tend to focus on learning local node representations, making it difficult to effectively capture structural relationships between node pairs. Furthermore, excessive reliance on local neighborhood information can lead to over-smoothing. Prior studies have shown that introducing global structural encoding can partially alleviate this issue.
To address these limitations, we propose a \textbf{Community-Enhanced Link Prediction (CELP)} framework that incorporates community structure to jointly model local and global graph topology. Specifically, CELP enhances the graph via community-aware, confidence-guided edge completion and pruning, while integrating multi-scale structural features to achieve more accurate link prediction. Experimental results across multiple benchmark datasets demonstrate that CELP achieves superior performance, validating the crucial role of community structure in improving link prediction accuracy.
\end{abstract}



\begin{keyword}


Graph Neural Networks \sep 
Link Prediction \sep 
Community Detection \sep 
Structure Enhancement
\end{keyword}

\end{frontmatter}



\section{Introduction}
Link prediction~\cite{zhang2018link, ying2018graph}, as one of the fundamental tasks in graph analysis~\cite{scarselli2008graph, xu2018powerful}, plays a crucial role in applications such as social recommendation \cite{he2020lightgcn, hu2018leveraging}, biomolecular interaction prediction \cite{ioannidis2020few, wang2022molecular}, and chain optimization \cite{yang2021financial}. In recent years, research in this area has converged into two primary paradigms: edge-centric methods and node-centric methods~\cite{kumar2020link, wang2024optimizing}.
Edge-centric methods \cite{zhang2018link} approach link prediction by explicitly modeling the structural relationship between node pairs. These methods typically construct enclosing subgraphs around candidate edges and classify them using graph neural networks, effectively capturing structural patterns such as common neighbors or closed triangles. While this strategy yields high discriminative power in local structures, it often suffers from high computational overhead, especially when applied to large-scale graphs.
Node-centric methods \cite{wang2023neural}, in contrast, focus on learning meaningful node embeddings that reflect both local attributes and topological information. The link likelihood is then estimated based on the similarity or proximity of the learned node representations. These methods are built upon the assumption that nodes with similar features or those embedded in analogous structural contexts are more likely to be connected. By decoupling the prediction process from subgraph construction, node-centric approaches offer superior scalability and efficiency, making them suitable for sparse and large graphs.

Although these link prediction methods demonstrate strong expressive power, recent studies have revealed that node representations in GNNs tend to converge after multiple propagation layers, leading to diminished discriminability—a phenomenon known as over-smoothing \cite{chen2020measuring}. To address this issue, previous approaches \cite{zheng2021cold, vaswani2017attention} have drawn inspiration from the positional encoding mechanism in Transformers, incorporating structural encodings combined with original features to enhance the diversity of node representations and thereby mitigate over-smoothing.

However, such methods still suffer from limitations in modeling the global structural information of graphs. Conventional node encoding schemes, such as random walk-based or degree-based approaches, primarily rely on stochastic mechanisms to preserve representation diversity, yet fail to explicitly capture a node’s functional role within the overall graph structure \cite{ribeiro2017struc2vec}. This often compromises the interpretability and structural robustness of the learned embeddings, especially in real-world graphs characterized by ambiguous semantic boundaries \cite{donnat2018learning}.

\begin{figure}[t]
\centering
\includegraphics[width=0.75\textwidth]{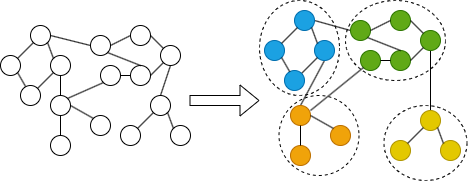}
\caption{\textbf{Illustrative Example of the Community Detection Algorithm.} 
By applying a community detection algorithm, a complex graph can be partitioned into different communities, where nodes within the same community are more densely connected, while connections with nodes from different communities are relatively sparse.}
\label{fig:community_detection}
\end{figure}

Therefore, we seek a representation learning paradigm that explicitly reflects global structural features of the graph. Theoretically, the task of community detection \cite{girvan2002community, blondel2008fast} offers natural advantages in both structural interpretability and hierarchical modeling. As illustrated in Figure~\ref{fig:community_detection}, a graph can be decomposed into multiple communities with dense intra-community connections and relatively sparse inter-community links. Within each community, nodes exhibit heterogeneous structural roles, where certain nodes act as representative or central figures,
while others primarily serve peripheral or bridging functions across communities. Such hierarchical organization offers a coarse-to-fine view of graph topology, which is difficult to capture using purely local neighborhood aggregation.
On one hand, community structures reveal densely connected subregions within a graph, often corresponding to functional units such as interest groups in social networks or departmental clusters in organizational graphs. Among various frameworks, Newman’s modularity maximization \cite{newman2006modularity} stands out as a widely adopted criterion, which evaluates community quality by measuring the deviation of intra-community edge density from that expected in a random network. On the other hand, community structure also facilitates a hierarchical representation mechanism. In social networks, individuals within a group often exhibit varying degrees of importance. For example, highly connected or mediating users can be seen as central figures within the community, while nodes linking multiple groups embody bridging roles across the global network. This perspective enables a more holistic understanding of structural dependencies among node pairs.

Despite the representational power of community structure, current mainstream link prediction methods have yet to incorporate such information. This leads to suboptimal performance in capturing cross-community links and dealing with structurally ambiguous regions. Moreover, existing approaches often fail to integrate community membership with node-pair representations, thus underutilizing the potential of community structure to enhance edge-level neighborhood semantics.

To address these limitations, we propose a novel Community-Enhanced Link Prediction (CELP) framework that leverages community structure to jointly refine graph topology and learn edge representations. By incorporating community-aware structural priors into both graph enhancement and representation learning, CELP effectively integrates global and local structural information for link prediction. We summarize our main contributions as follows:

\begin{itemize}
    \item We propose a novel community-enhanced link prediction framework, CELP, which leverages community structure as a global structural prior to refine the
    observed graph topology. Specifically, CELP performs community-aware, confidence-guided edge completion and pruning to mitigate missing or noisy connections in the graph.

    \item We design a multi-scale, community-aware edge representation scheme that integrates local neighborhood information, path-based structural features, and
    community-level relational context. By explicitly incorporating community assignments and inter-community structural distances into node-pair encoding, CELP captures both fine-grained local patterns and global topological dependencies.

    \item We conduct extensive experiments on multiple real-world datasets with varying scales, demonstrating that CELP consistently outperforms state-of-the-art baselines in link prediction tasks.
\end{itemize}

\section{Related Work}
Link prediction, a fundamental task in graph analysis, aims to infer missing or future connections between nodes, with broad applications ranging from social network recommendation \cite{he2020lightgcn, hu2018leveraging} to biological interaction prediction \cite{ioannidis2020few, wang2022molecular}. Traditional heuristic approaches \cite{barabasi1999emergence, zhou2009predicting} primarily leverage structural similarity metrics derived from local topological patterns. A seminal work by Adamic and Adar \cite{adamic2003friends} introduced a weighted common-neighbor index, where the contribution of each shared neighbor is inversely weighted by its degree, effectively capturing the intuition that connections via low-degree neighbors are more informative. This builds upon earlier neighborhood-based methods such as the Jaccard index \cite{wu2019adversarial}, which quantifies overlap in immediate neighbors, and the preferential attachment score, which assumes higher-degree nodes are more likely to form new links. Other path-based heuristics, including the Katz index \cite{martinez2016survey}, extend this paradigm by incorporating higher-order proximity through bounded-length paths. While these methods are computationally efficient and interpretable, their reliance on handcrafted structural features limits their capacity to model complex nonlinear dependencies in real-world graphs.

With the advancement of deep learning techniques, researchers have increasingly incorporated graph neural networks (GNNs) into link prediction tasks. Methods like GraphSAGE \cite{hamilton2017inductive} have demonstrated significant improvements by sampling and aggregating information from neighboring nodes, enabling effective handling of large-scale graphs while achieving superior prediction performance. These approaches learn latent representations for each node and utilize these embeddings to predict potential edges between node pairs. GNNs are particularly well-suited for link prediction as they can effectively capture both structural patterns and complex relational dependencies within graphs. Popular architectures including Graph Convolutional Networks (GCNs) \cite{kipf2016semi} and Graph Attention Networks (GATs) \cite{velivckovic2017graph} have been successfully adapted for this task, employing either neighborhood convolution operations or self-attention mechanisms to learn expressive node representations that significantly enhance link prediction accuracy.

These methods can be broadly categorized into two paradigms \cite{wang2024optimizing}: edge-centric methods and node-centric methods. Edge-centric approaches typically construct local structural subgraphs around candidate edges for classification, while node-centric methods focus on learning high-quality node representations to better capture potential connectivity patterns.

Edge-Centric Methods primarily employ subgraph classification strategies. SEAL  \cite{zhang2018link} reformulated link prediction as a graph classification problem by extracting enclosing subgraphs centered on node pairs and processing them with deep graph neural networks. While effectively capturing local structural patterns, these methods face computational bottlenecks when scaling to large graphs. SHFF \cite{liu2020feature} introduced hierarchical feature fusion to aggregate node features across subgraph levels, yielding more discriminative subgraph representations. LGLP \cite{cai2021line}, , inspired by SEAL's innovative approach of transforming link prediction tasks into graph classification problems, transformed edge information into node features via graph conversion, applying node classification algorithms instead. Despite enhanced structural expressiveness, these approaches still incur substantial computational overhead during subgraph generation and encoding.

Node-Centric Methods diverge by concentrating on node representation learning rather than subgraph construction. Neo-GNN \cite{yun2021neo} pioneered this direction with dual feature extractors for nodes and edges, comprehensively modeling structural relationships. BUDDY \cite{chamberlain2022graph} augmented node pairs with neighbor feature extremum matrices, while NCNC \cite{wang2023neural} systematically exploited common neighbor features to enhance pairwise modeling. For efficient neighborhood encoding, several innovations emerged: Bloom Signatures \cite{zhang2024learning} compressed neighbor information using compact signature structures; MPLP \cite{dong2024pure} employed quasi-orthogonal vectors for link representation, balancing discriminability with efficiency; LPFormer \cite{shomer2024lpformer} combined MPNNs with pairwise encoding mechanisms for optimal expressivity-efficiency trade-offs. These methods demonstrate superior scalability, particularly for large-scale and sparse graphs.

Although existing link prediction models have demonstrated strong performance across various benchmark datasets, they still face notable limitations: some struggle to capture the global structural patterns of the graph, while others rely heavily on complex subgraph sampling procedures. In this study, we draw inspiration from Mao et al.~\cite{mao2023revisiting}, who systematically identified three key factors affecting link prediction performance: local structural similarity, global structural similarity, and node feature similarity. Based on these insights, we propose CELP, a method that incorporates community structure through graph partitioning to better model global information. CELP leverages community-level priors to complete the graph structure, thereby enhancing local structural cues. Additionally, it integrates community membership with neighborhood features to enrich edge representations, ultimately improving the prediction accuracy.

\begin{figure*}[t]
\centering
\includegraphics[width=1\textwidth]{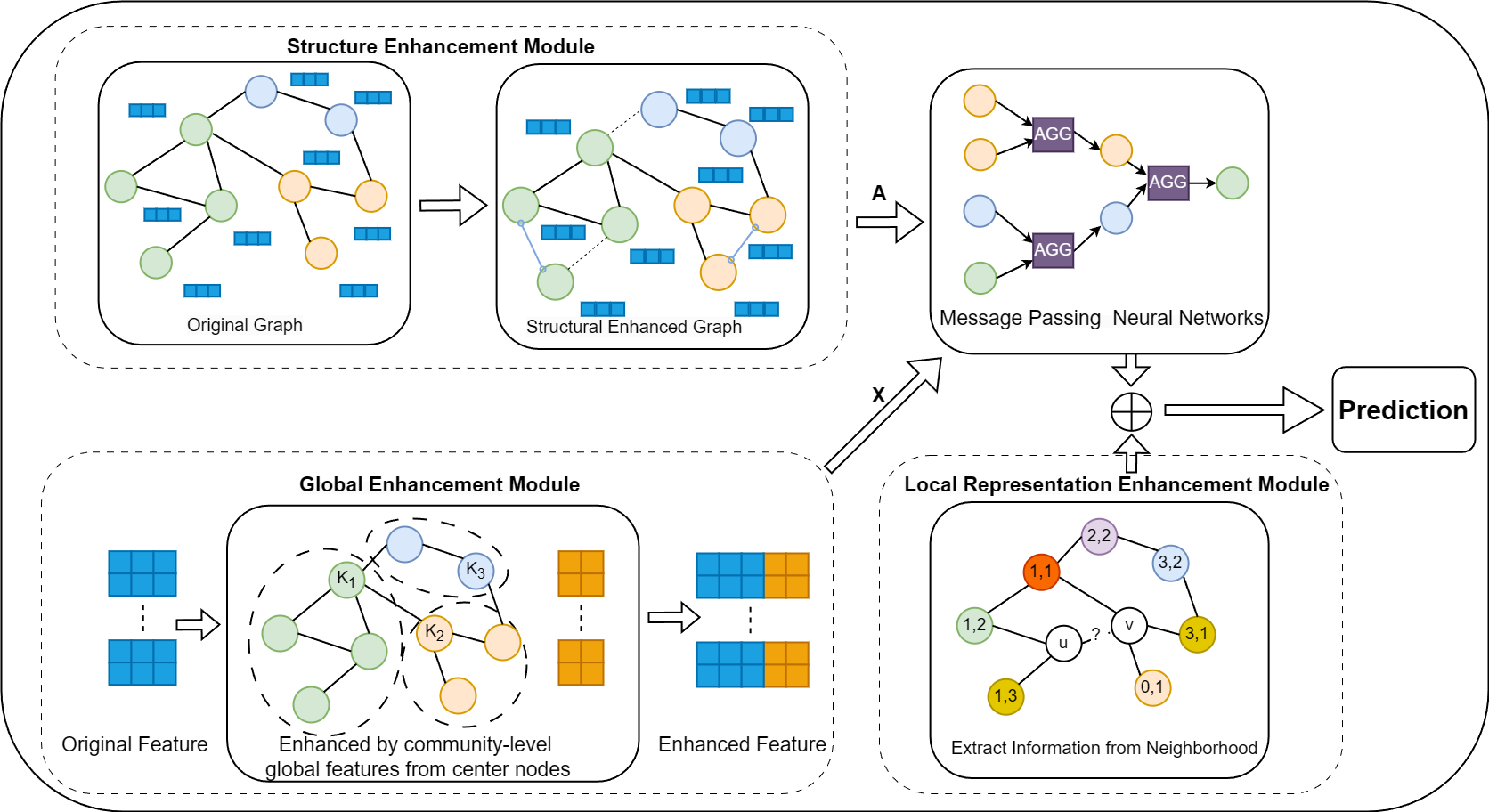}
\caption{\textbf{Framework of CELP}. Community detection is first performed on the original graph, and the PageRank centrality is utilized to identify the central node within each community, which serves as the foundation for constructing global node representations. Next, prior probabilities are introduced to enhance and complete the graph structure, effectively mitigating the issue of graph incompleteness. Finally, by integrating the local neighborhood features, path information, and cross-community collaborative features of target node pairs, a more effective edge representation is constructed, thereby improving the performance of link prediction.}
\label{fig:framework}
\end{figure*}

\section{Preliminaries}
Given an undirected graph \( G = (V, E, X) \) with node features, where \( V = \{v_1, v_2, \dots, v_N\} \) denotes the set of \( N \) nodes and \( E \subseteq V \times V \) is the set of observed edges, each node \( v_i \in V \) is associated with a feature vector \( x_i \in \mathbb{R}^d \). The feature matrix \( X \in \mathbb{R}^{N \times d} \) stacks all node feature vectors row-wise, where \( d \) is the feature dimension. The graph structure is encoded by the adjacency matrix \( A \in \{0, 1\}^{N \times N} \), where \( A_{ij} = 1 \) indicates an edge between nodes \( i \) and \( j \), and 0 otherwise. The immediate neighborhood of a node \( v \), denoted by \( \mathcal{N}_v \), is the set of nodes adjacent to \( v \). The degree of node \( v \) is then given by \( d_v = |\mathcal{N}_v| \). 

Link prediction aims to estimate the existence likelihood of edges between pairs of nodes that are not observed in the graph. 
Given the graph \( G = (V, E, X) \), a graph neural network is first employed to learn node representations 
\( \mathbf{H} = \{ \mathbf{h}_v \mid v \in V \} \), where \( \mathbf{h}_v \in \mathbb{R}^d \) denotes the embedding of node \( v \).
For a pair of nodes \( (u, v) \), an edge representation \( \mathbf{e}_{uv} \) is then constructed based on their node embeddings,
for example via element-wise product or a multilayer perceptron:
\begin{equation}
\mathbf{e}_{uv} = \phi(\mathbf{h}_u, \mathbf{h}_v),
\end{equation}
where \( \phi(\cdot) \) denotes a pairwise feature mapping function.
Finally, a prediction function is applied to estimate the probability of an edge between nodes \( u \) and \( v \).

We note that most community detection algorithms rely solely on the
structural information of a graph.
Accordingly, we perform community detection on the structural graph
$G_{\text{struct}} = (V, E)$ to partition the node set into $K$ disjoint
communities:
\begin{equation}
\mathcal{C} = \left\{ \mathcal{C}_1, \mathcal{C}_2, \ldots, \mathcal{C}_K \right\},
\end{equation}
where each community $\mathcal{C}_k \subseteq V$ is a set of nodes,
$\bigcup_{k=1}^{K} \mathcal{C}_k = V$, and
$\mathcal{C}_i \cap \mathcal{C}_j = \emptyset$ for $i \neq j$.

For notational clarity, we use $\mathcal{C}_k$ to denote the node set of the $k$-th community, and $s_u \in \{1,\ldots,K\}$ to denote the community assignment of node $u \in V$, i.e., $u \in \mathcal{C}_{s_u}$.
Furthermore, we denote by $r_k$ the representative center node of community $\mathcal{C}_k$, and define $c_u := r_{s_u}$ as the center node of the community to which node $u$ belongs.

\section{Method}
In this section, we present our proposed Community-Enhanced Link Prediction (CELP) framework, which systematically leverages community structure information to enhance link prediction performance. As illustrated in Figure~\ref{fig:framework}, the CELP framework incorporates three key innovations: Global Enhancement Module, Structure Enhancement Module, and Local Representation Enhancement Module.

\subsection{Global Enhancement Module}
To address the oversmoothing problem in graph neural networks (GNNs), we introduce a novel community-aware structural encoding strategy. Unlike prior approaches that learn structural embeddings implicitly, our method constructs explicit, interpretable, and efficient global structure representations based on graph topology. 

Community detection is a core task in the field of graph mining, aiming to partition the vertex set into several disjoint subsets such that nodes within the same community are densely connected, while connections between different communities are sparse. This structure reflects the underlying organization or functional modules in a graph and has broad practical significance in applications such as social networks, recommendation systems, and biological networks.
In this study, we employ the FluidC \cite{pares2018fluid} algorithm as our community detection method. FluidC is an efficient graph clustering algorithm inspired by the simulation of fluid diffusion. The core of FluidC lies in its community update rule. Specifically, in each iteration, the algorithm traverses all vertices in a random order. For each vertex $v$, it selects the most suitable target community by aggregating evidence from its ego network. 
The update rule of FluidC can be expressed as:
\begin{equation} 
s_{v}^{\prime}
= \arg \max_{k \in \{1,\ldots,K\}}
\sum_{w \in \{v\} \cup \mathcal{N}_v}
\omega(s_w)\cdot \mathbf{1}_{[s_w = k]}
\end{equation}
Here, $\mathcal{N}_v$ denotes the set of neighbors of vertex $v$, and $\{v\} \cup \mathcal{N}_v$ represents the ego network of $v$, including the node itself and its immediate neighbors.
The weighting term $\omega(s_w)$ is defined as:
\begin{equation}
\omega(s_w) = \frac{1}{|\mathcal{C}_{s_w}|}
\end{equation}
which serves as a community size normalization factor to prevent large communities from dominating the update process.
The indicator function $\mathbf{1}_{[s_w = k]}$ equals 1 if vertex $w$ belongs to community $k$, and 0 otherwise.

\begin{table}[t]
\centering
\renewcommand\arraystretch{1.3}

\caption{Results with different indicators to determine the influence of nodes in the network by HR@100. The best result is \textbf{bolded}, and the second best is marked with $^{*}$.}
\label{tab:centrality}

\begin{tabular}{l c c c}
\hline
 & Cora & Citeseer & Pubmed \\
\hline
Degree centrality 
& 91.67 $\pm$ 0.92 
& 92.24 $\pm$ 1.62 
& 82.70 $\pm$ 1.76 \\

Betweenness centrality 
& 91.88$^{*}$ $\pm$ 0.74 
& \textbf{94.20 $\pm$ 1.98} 
& 83.12$^{*}$ $\pm$ 1.96 \\

Closeness centrality 
& 91.24 $\pm$ 1.36 
& 93.46 $\pm$ 1.52 
& 82.22 $\pm$ 2.12 \\

PageRank centrality 
& \textbf{92.41 $\pm$ 1.06} 
& 93.70$^{*}$ $\pm$ 1.83 
& \textbf{83.41 $\pm$ 1.64} \\
\hline
\end{tabular}

\end{table}

After obtaining the community partition results, we compute the most central node within each community to extract global structural information that reflects the importance of nodes in the entire graph. Based on the comparative analysis in Table \ref{tab:centrality} among four centrality measures: Degree Centrality, Betweenness Centrality \cite{brandes2001faster}, Closeness Centrality \cite{chen2023normalized}, and PageRank Centrality \cite{page1999pagerank} in terms of HR@100 performance, we observe that PageRank Centrality consistently achieves the best or second-best results across all three datasets. Although Betweenness Centrality also performs well, its high computational complexity may become a bottleneck when applied to large-scale graphs. Therefore, considering both predictive performance and computational efficiency, we adopt PageRank Centrality as the default method for computing node importance.

Specifically, for each community \( C_k \), we compute the importance scores of all nodes using the PageRank algorithm and select the node with the highest PageRank score as the representative center of that community. PageRank is originally defined on directed graphs. Since the graph considered in this work is undirected, we follow the common practice of converting each undirected edge into two reciprocal directed edges. Under this setting, the in-degree 
and out-degree of each node are identical. The PageRank score captures not only the local connectivity of a node but also the global topological structure of the entire network, enabling effective identification of representative hub nodes. The PageRank score of each node is computed as follows:
\begin{equation} 
PR(v) = \frac{1 - \alpha}{N} + \alpha \sum_{u \in \mathcal{N}_v^{-}} \frac{PR(u)}{d^{+}(u)}
\end{equation}
where \( PR(v) \) denotes the PageRank score of node \( v \), 
\( \alpha \) is the damping factor (typically set to 0.85), and \( N \) is the total number of nodes in the graph.
\( \mathcal{N}_v^{-} \) denotes the set of in-neighbors of \( v \), which coincides with its neighborhood in the undirected graph, and \( d^{+}(u) \) denotes the out-degree of node \( u \).

To capture the relative structural position of each node in the global network, we construct a structural encoding vector \(\mathbf{str}_v \in \mathbb{R}^K\) for each node \(v \in V\), where the \(k\)-th element represents the shortest path distance from node \(v\) to the \(k\)-th community center \(r_k\), defined as:
\begin{equation}
\mathbf{str}_v(k) = SPD(v, r_k)
\end{equation}
where \(SPD(v, r_k)\) denotes the shortest path length from node \(v\) to community center \(r_k\) in the graph \(G\). The resulting vector captures the relative positions of node \(v\) with respect to multiple community centers, providing a compact representation of its structural role across communities.

Finally, we concatenate the normalized structural encoding vector \(\mathbf{str}_v\) with the original node features \(\mathbf{x}_v\) to obtain the enhanced input representation of node $v$:

\begin{equation}
\hat{\mathbf{x}}_v = \text{concat}(\mathbf{x}_v, \mathbf{str}_v)
\end{equation}

This joint representation integrates both the intrinsic attribute information of the node and its structural position within the entire graph. By incorporating structural encodings, the model is able to preserve node-level distinctions even after multiple rounds of message passing \cite{zheng2021cold}. As a result, this strategy effectively alleviates the over-smoothing issue and enhances the model's ability to distinguish nodes with different structural roles.

\subsection{Structure Enhancement Module}
Given that graph incompleteness is prevalent in link prediction tasks, it may cause a distribution shift between the training and testing sets, as well as the loss of crucial structural information such as common neighbors, ultimately degrading model performance. To address this issue, we propose a structure enhancement strategy based on prior probabilities: We identify a subset of non-adjacent node pairs that are more likely to correspond to missing or unobserved links, and treat them as high-confidence missing edges. These edges are further regarded as potential positive edges and incorporated back into the adjacency matrix. This approach helps compensate for missing structural information, reduce the model’s sensitivity to noise, and improve link prediction accuracy on incomplete graphs.

However, considering all non-adjacent node pairs as candidate edges would result in a candidate space of size \( O(N^2) \), which is computationally prohibitive for large-scale graphs. Inspired by \cite{kim2025accurate, wang2024optimizing}, we therefore construct a reduced candidate set. Unlike previous approaches that filter candidates based on node degree, we adopt a filtering strategy based on node importance scores derived from the PageRank algorithm. Specifically, we retain only those candidate edges for which at least one endpoint belongs to the top-\(M\) nodes with the highest PageRank scores, denoted as \( V_{\text{top}M} \). In our experiments, we set \( M = 100 \). The final candidate edge set is defined as:
\begin{equation}
E_{\text{cand}} = \{ (u, v) \mid u \in V_{\text{top}M} \ \text{or} \ v \in V_{\text{top}M} \}
\end{equation}

Building upon the initial candidate edge set, we introduce an additional filtering condition: for a candidate node pair $(u, v)$, both nodes must belong to the same community, i.e., \( s_u = s_v \). The rationale behind this design lies in the fact that community structures reflect local structural similarity and functional proximity among nodes in a graph. Nodes within the same community are more likely to be semantically related and structurally connected. By restricting candidate edges to intra-community pairs, we not only reduce unnecessary computational overhead but also enhance the precision of the candidate edge set, lowering the risk of introducing noisy or unlikely connections. Moreover, this community-aware edge generation strategy better aligns with realistic patterns observed in graph-structured data and can significantly improve the performance of subsequent link prediction models.We define the refined candidate edge set with community constraints as:
\begin{equation}
E_{\text{cand2}} = \{ (u, v) \in E_{\text{cand}} \mid s_u = s_v \}
\end{equation}

After obtaining the refined candidate edge set, we design a confidence-based filtering mechanism to complete missing edges in the graph. Since improper edge addition may introduce low-quality negative samples, which can degrade the performance of the link prediction model, it is essential to adopt a rigorous selection strategy. Specifically, we first pre-train a link prediction model $F(u,v \mid G)$ to estimate the connection probability $p_{u,v} = F(u,v \mid G)$ for any node pair $(u,v)$ based on the current graph structure $G$. Then, for each candidate edge in $E_{\text{cand2}}$, we select the top-$\gamma$ proportion of edges with the highest predicted probabilities, resulting in the final set of high-confidence added edges:
\begin{equation}
E_\text{add} = \text{Top-}\gamma \left( \{ (u, v) \in E_{\text{cand2}} \mid p_{u,v} \} \right)
\end{equation}
where $\gamma$ is the proportion parameter that controls the number of edges to be added.

In addition to completing missing edges, we further perform a confidence-based edge pruning step to remove potentially noisy edges from the original training graph. Specifically, we identify the bottom-\(\eta\) proportion of existing edges in the original edge set \( E \) according to their predicted probabilities and regard them as unreliable connections:
\begin{equation}
E_{\text{rm}} = \text{Bottom-}\eta \left( \{ (u, v) \in E \mid p_{u,v} \} \right),
\end{equation}
where \( \eta \) controls the proportion of edges to be removed during pruning.

Finally, we update the graph structure by removing low-confidence edges and incorporating the newly added high-confidence edges:
\begin{equation}
E_{\text{new}} = (E \setminus E_{\text{rm}}) \cup E_{\text{add}}.
\end{equation}

Based on the updated edge set \( E_{\text{new}} \), we construct a new adjacency matrix \( \hat{A} \) for subsequent GNN training and representation learning. This dual strategy --- removing unreliable edges while completing missing ones --- not only strengthens the structural integrity of the graph, but also mitigates the impact of mislabeled or noisy links. It thus enhances both the robustness and generalization ability of the model in real-world incomplete or imperfect graphs.

\subsection{Local Representation Enhancement Module}

While Message Passing Neural Networks (MPNNs) have demonstrated remarkable success in node-level representation learning, they exhibit significant limitations in encoding joint structural relationships between candidate node pairs for link prediction tasks. For instance, key structural metrics like common neighbor counts cannot be effectively captured through standard local node aggregation. Recent studies have shown that combining MPNN-generated node embeddings with explicit pairwise structural relationship encoding can substantially improve link prediction performance. Inspired by these findings \cite{yun2021neo, chamberlain2022graph, wang2023neural}, we propose a triple-relational edge representation enhancement module that integrates neighborhood features, path-based features, and community-aware features to enrich both the semantic and structural expressiveness of edge representations.

\subsubsection{Neighborhood Features} 
To more comprehensively capture the local topological structure of node pairs, we introduce Distance Encoding (DE) \cite{li2020distance} as a representation enhancement strategy. DE is a partitioning method based on the Shortest Path Distance (SPD), where each node $k$ is assigned to the partition $DE(p, q)$ corresponding to its distances from $u$ and $v$, i.e., $DE(p, q) = \{ k \mid \text{SPD}(u, k) = p, \ \text{SPD}(v, k) = q \}$.

Inspired by the MPLP \cite{dong2024pure} approach, we improve DE in two key aspects: On the one hand, we perform a representation transformation by repurposing DE from a node-level labeling tool to a structural feature representation mechanism tailored for node pairs in link prediction tasks; On the other hand, we introduce computational optimization by mapping the DE coordinates into a quasi-orthogonal (QO) vector space \cite{kainen2020quasiorthogonal, nunes2023dothash}, which significantly reduces the encoding complexity while retaining informative topological signals from the $k$-hop neighborhood. 
This design enables each node pair to acquire fine-grained topological awareness, thus enhancing the model’s expressive capacity in structurally intricate scenarios. This process is formalized as:
\begin{equation}
\mathbf{f}^{p,q}(u,v) = \mathbf{\eta}^{p}_u \cdot \mathbf{\eta}^{q}_v
\end{equation}
where $\mathbf{\eta}^{p}_u \in \mathbb{R}^d$ and $\mathbf{\eta}^{q}_v \in \mathbb{R}^d$ denote the quasi-orthogonal vector representations of the $p$-hop and $q$-hop neighborhoods of nodes $u$ and $v$, respectively, and $\cdot$ denotes the inner product. This formulation captures the soft overlap between the multi-hop neighborhoods by measuring the similarity of their propagated QO embeddings.

\subsubsection{Path-Based Features}
Previous studies have shown that multi-hop path information plays a crucial role in modeling the structural proximity between node pairs~\cite{zhang2023page, lv2024path}. To capture such higher-order structural semantics, we first introduce the theoretical formulation of Personalized PageRank (PPR):
\begin{equation}
\mathbf{\Pi} = \beta \big(I - (1-\beta)D^{-1}A \big)^{-1},
\end{equation}
where $\beta$ is the restart probability and $D^{-1}A$ denotes the random-walk transition matrix. This formulation characterizes global structural proximity by aggregating paths of all lengths with exponentially decayed weights. Note that the closed-form expression is used only for theoretical exposition and is not computed explicitly in practice.

To enable efficient path-based feature modeling, we adopt a truncated PPR approximation that avoids explicit matrix inversion by limiting the number of hops:
\begin{equation}
\mathbf{p}^{r}_u = \sum_{k=0}^{r} \beta (1 - \beta)^k A^k \boldsymbol{\eta}_u,
\end{equation}
where $\boldsymbol{\eta}_u$ is the quasi-orthogonal vector associated with node $u$, and $r$ denotes the maximum hop distance. This truncated expansion provides an efficient approximation of PPR while retaining the ability to capture multi-hop structural dependencies.

For a node pair $(u, v)$, we define the path-based structural feature as the element-wise interaction between their truncated PPR vectors:
\begin{equation}
\mathbf{g}^{r}(u, v) = \mathbf{p}^{r}_u \odot \mathbf{p}^{r}_v.
\end{equation}
where $\odot$ denotes the Hadamard product.

This path-based representation enables the model to capture both multi-hop connectivity and overlapping neighborhood structures with exponential decay. In particular, it not only distinguishes between strong and weak indirect connections by integrating higher-order paths, but also incorporates both local and global structural semantics into node pair modeling, thereby providing a more comprehensive structural characterization.

Similarly to preceding works \cite{yin2022algorithm, wang2023neural, dong2024pure}, we remove target link to alleviate potential distributional bias between the training and testing sets. In the testing phase, the model is required to predict links that are inherently missing from the graph. However, such missing links could either be actual negatives or merely absent due to incomplete graph construction. If these target links are retained during training, the model might exploit such ``shortcut'' connections, leading to overfitting on the structural patterns present in the training graph.

For example, consider a target node pair $(u, v)$, and suppose there exists a neighbor node $k$ that is one hop away from $u$, but multiple hops from $v$. If the link between $u$ and $v$ is preserved during training, node $k$ may use this direct connection as a shortcut to reach $v$, thereby artificially reducing the shortest path distance. This results in compressed path features for positive samples and creates a structural distribution shift between training and testing data.

Similarly, path counts may also be distorted. Retaining the target link can introduce additional shortcut-based paths into the statistics. For instance, a 3-hop path may actually be composed of a 2-hop path plus the direct target link, leading to inflated path counts. This undermines the accuracy of path-based structural modeling and further increases the mismatch between training and testing conditions.

To address these issues, we explicitly remove target links during training. This ensures consistent and unbiased neighborhood modeling, and promotes more faithful and robust structural representation learning that generalizes better during testing.

\subsubsection{Community-Aware Features}
Community structure is a fundamental aspect of real-world networks, significantly influencing node connectivity patterns. Empirical studies have consistently shown that nodes within the same community are more likely to be connected due to shared properties, functions, or roles, while inter-community connections are modulated by the structural distance and interaction intensity between communities. Neglecting such organization may lead to suboptimal modeling of network structure in link prediction tasks.

Therefore, we propose a community-aware feature design that explicitly encodes the relational context between node pairs. For a given node pair $(u, v)$, we first identify their respective community assignments $s_u$ and $s_v$, and extract community-level features by jointly considering their community identities and the structural relationship between their corresponding communities.
Specifically, we incorporate the community assignments of both nodes, as well as the shortest path distance between the center nodes of their associated communities, to capture inter-community structural proximity.
We formally define the community-aware feature for a node pair $(u, v)$ as:
\begin{equation}
\mathbf{com}(u, v) = \text{concat}\left(
s_u,\; s_v,\; \mathrm{SPD}(c_u, c_v) \right). \end{equation}
where $s_u$ and $s_v$ denote the community assignments of nodes $u$ and $v$, respectively. $c_u := r_{s_u}$ and $c_v := r_{s_v}$ denote the center nodes of the communities to which $u$ and $v$ belong. $\mathrm{SPD}(c_u, c_v)$ represents the shortest path distance between the two community center nodes.
\subsection{Final Representation}

\begin{table*}[t]
\centering
\renewcommand\arraystretch{1.3}
\caption{Results on link prediction benchmarks. The format is average score $\pm$ standard deviation. 
In each column, the boldfaced score denotes the best result, and the second best is marked with $^{*}$.}
\label{tab:link_prediction_results}

\resizebox{\columnwidth}{!}{
\begin{tabular}{l c c c c c c}
\hline
 & Cora & Citeseer & Pubmed & Photo & Computers & Collab \\
\hline
Metric 
& HR@100 & HR@100 & HR@100 & HR@50 & HR@50 & HR@50 \\
\hline
CN        
& 33.92 $\pm$ 0.46 
& 29.79 $\pm$ 0.90 
& 23.13 $\pm$ 0.15 
& 29.33 $\pm$ 2.74 
& 21.95 $\pm$ 2.00 
& 61.37 $\pm$ 0.00 \\

AA        
& 39.85 $\pm$ 1.34 
& 35.19 $\pm$ 1.33 
& 27.38 $\pm$ 0.11 
& 37.35 $\pm$ 2.65 
& 26.96 $\pm$ 2.08 
& 64.35 $\pm$ 0.00 \\

RA        
& 41.07 $\pm$ 0.48 
& 33.56 $\pm$ 0.17 
& 27.03 $\pm$ 0.35 
& 40.77 $\pm$ 3.41 
& 28.05 $\pm$ 1.59 
& 64.00 $\pm$ 0.00 \\
\hline
GCN       
& 66.79 $\pm$ 1.65 
& 67.08 $\pm$ 2.94 
& 53.02 $\pm$ 1.31 
& 28.14 $\pm$ 7.81 
& 22.95 $\pm$ 10.58 
& 35.53 $\pm$ 2.39 \\

SAGE      
& 55.02 $\pm$ 4.03 
& 57.01 $\pm$ 3.74 
& 39.66 $\pm$ 2.70 
& 46.01 $\pm$ 1.83 
& 33.79 $\pm$ 3.11 
& 36.82 $\pm$ 7.41 \\
\hline
SEAL      
& 81.71 $\pm$ 1.30 
& 83.89 $\pm$ 2.15 
& 75.54 $\pm$ 1.32 
& 46.08 $\pm$ 3.27 
& 30.43 $\pm$ 2.07 
& 64.74 $\pm$ 0.43 \\

Neo-GNN   
& 80.42 $\pm$ 1.31 
& 84.67 $\pm$ 2.16 
& 73.93 $\pm$ 1.19 
& 44.83 $\pm$ 3.23 
& 22.76 $\pm$ 3.07 
& 57.52 $\pm$ 0.37 \\

BUDDY     
& 88.00 $\pm$ 0.44 
& 92.93 $\pm$ 0.27 
& 74.10 $\pm$ 0.78 
& 43.51 $\pm$ 2.37 
& 29.01 $\pm$ 2.66 
& 65.94 $\pm$ 0.58 \\

MPLP      
& 87.94 $\pm$ 1.22 
& 88.60 $\pm$ 2.18 
& 79.34 $\pm$ 1.23 
& 58.08$^{*}$ $\pm$ 3.68
& 43.47$^{*}$ $\pm$ 3.61 
& 67.05$^{*}$ $\pm$ 0.51 \\

NCNC      
& 89.65$^{*}$ $\pm$ 1.36
& 93.47$^{*}$ $\pm$ 0.95
& 81.29$^{*}$ $\pm$ 0.95
& 47.98 $\pm$ 2.36 
& 36.48 $\pm$ 4.16 
& 66.61 $\pm$ 0.71 \\
\hline
\textbf{CELP} 
& \textbf{93.34 $\pm$ 0.54} 
& \textbf{95.41 $\pm$ 0.69} 
& \textbf{84.11 $\pm$ 1.57} 
& \textbf{58.87 $\pm$ 2.98} 
& \textbf{43.81 $\pm$ 2.95} 
& \textbf{67.37 $\pm$ 0.56} \\
\hline
\end{tabular}
}
\end{table*}

To capture multi-scale distance-based structural information, we aggregate distance-based features computed at different hop combinations. Specifically, given a predefined hop set $\mathcal{H}_f$, we define the distance-based feature vector for a node pair $(u,v)$ as:
\begin{equation}
\mathbf{f}(u,v)
=
\text{concat}\left(
\left\{
\mathbf{f}^{p,q}(u,v)
\mid
p,q \in \mathcal{H}_f
\right\}
\right).
\end{equation}

Similarly, to encode path-based structural information at different scales, we consider truncated PPR representations with varying hop radii. Let $\mathcal{H}_g$ denote the set of hop values used for path-based modeling. The aggregated path-based feature for a node pair $(u,v)$ is defined as:
\begin{equation}
\mathbf{g}(u,v)
=
\text{concat}\left(
\left\{
\mathbf{g}^{r}(u,v)
\mid
r \in \mathcal{H}_g
\right\}
\right).
\end{equation}

Based on the above structural features, including the distance-based feature $\mathbf{f}(u,v)$, the path-based feature $\mathbf{g}(u,v)$, and the community-aware feature $\mathbf{com}(u,v)$, we construct the final representation for a node pair $(u,v)$ by integrating them with the learned
node embeddings from a GNN.

Specifically, we first compute the interaction between the node embeddings via inner product,
$\mathbf{h}_u^{\mathrm{GNN}} \cdot \mathbf{h}_v^{\mathrm{GNN}}$,
which captures semantic similarity in the embedding space. We then concatenate this interaction term with the aggregated structural features to obtain the final link representation:
\begin{equation}
\mathbf{h}(u, v)
=
\text{concat}\left(
\mathbf{h}_u^{\mathrm{GNN}} \cdot \mathbf{h}_v^{\mathrm{GNN}},
\mathbf{f}(u, v),
\mathbf{g}(u, v),
\mathbf{com}(u, v)
\right).
\end{equation}

This unified representation jointly encodes semantic, local, and global
structural information, and is subsequently fed into a classifier to
predict the existence of a link between nodes $(u,v)$.

\begin{table*}[t]
\centering
\renewcommand\arraystretch{1.3}
\caption{Ablation study results of CELP. ``GE'' denotes the Global Enhancement module, 
``SE'' denotes the Structure Enhancement module, and ``LE'' denotes the Local Representation Enhancement module.}
\label{tab:ablation_studies}
\resizebox{\columnwidth}{!}{
\begin{tabular}{l c c c c c c}
\hline
 & Cora & Citeseer & Pubmed & Photo & Computers & Collab \\
\hline
w/o LE and SE 
& 91.51 $\pm$ 1.10 
& 92.06 $\pm$ 2.42 
& 82.24 $\pm$ 1.47 
& 50.82 $\pm$ 3.15 
& 37.81 $\pm$ 3.74 
& 65.47 $\pm$ 0.80 \\

w/o GE and SE 
& 87.65 $\pm$ 1.64 
& 88.08 $\pm$ 2.18 
& 69.34 $\pm$ 1.23 
& 56.67 $\pm$ 2.29 
& 40.43 $\pm$ 4.22 
& 65.62 $\pm$ 0.64 \\

w/o GE and LE 
& 87.91 $\pm$ 1.63 
& 88.47 $\pm$ 2.32 
& 71.91 $\pm$ 1.66 
& 52.10 $\pm$ 2.44 
& 41.35 $\pm$ 2.95 
& 65.33 $\pm$ 1.01 \\
\hline
w/o SE 
& 91.77 $\pm$ 0.89 
& 93.92$^{*}$ $\pm$ 1.64
& 82.61 $\pm$ 1.51 
& 57.41 $\pm$ 3.74 
& 42.21 $\pm$ 3.70 
& 66.17 $\pm$ 0.45 \\

w/o LE 
& 91.93$^{*}$ $\pm$ 1.11 
& 93.34 $\pm$ 1.14 
& 83.09$^{*}$ $\pm$ 1.56
& 52.72 $\pm$ 3.33 
& 41.35 $\pm$ 2.95 
& 66.01 $\pm$ 0.57 \\

w/o GE 
& 88.34 $\pm$ 1.48 
& 92.88 $\pm$ 0.91 
& 78.59 $\pm$ 1.52 
& 58.07$^{*}$ $\pm$ 2.81
& 42.71$^{*}$ $\pm$ 3.12 
& 66.72$^{*}$ $\pm$ 0.45 \\
\hline
CELP 
& \textbf{93.34 $\pm$ 0.54} 
& \textbf{95.41 $\pm$ 0.69} 
& \textbf{84.11 $\pm$ 1.57} 
& \textbf{58.87 $\pm$ 2.98} 
& \textbf{43.81 $\pm$ 2.95} 
& \textbf{67.37 $\pm$ 0.56} \\
\hline
\end{tabular}
}

\end{table*}

\subsection{Loss Function}

Consistent with prevalent practices in prior link prediction work, we adopt the cross-entropy loss as the primary objective for model training. 
Let $y_{uv} \in \{0,1\}$ denote the ground-truth label indicating whether an edge exists between node pair $(u,v)$, and let $\hat{y}_{uv} \in [0,1]$ be the predicted probability of link existence. 
The supervised loss is formulated as:
\begin{equation}
\mathcal{L}_{\mathrm{CE}} = -\frac{1}{|\mathcal{E}_{\text{train}}|}\sum_{(u,v) \in \mathcal{E}_{\text{train}}} \Big[ y_{uv} \log \hat{y}_{uv} + (1-y_{uv}) \log (1-\hat{y}_{uv}) \Big],
\end{equation}
where $\mathcal{E}_{\text{train}}$ denotes the set of training edges and negative samples. 

In addition to the standard classification loss, we incorporate a structural contrastive loss $\mathcal{L}_{\mathrm{con}}$ that is similar in spirit to the method proposed in \cite{wang2025ngon}, but with a key difference. While~\cite{wang2025ngon} builds the constraint matrix based on ground-truth labels, we instead leverage community structure as a form of weak structural supervision.
Specifically, we construct a community assignment matrix
$\mathbf{S} \in \{0,1\}^{N \times K}$, where $K$ denotes the number of communities, and each row is a one-hot vector indicating the community membership of a node. The product $\mathbf{S}\mathbf{S}^{\top}$ yields a node-level structural
constraint matrix, where the $(i,j)$-th entry equals $1$ if nodes $i$ and $j$ belong to the same community, and $0$ otherwise.
Following the PPR-based constraint framework, we then construct the structural constraint matrix $\mathbf{M}$ as:
\begin{equation}
\mathbf{M} = \mathbf{S}\mathbf{S}^{\top} \odot \mathbf{\Pi},
\end{equation}
where $\odot$ denotes the Hadamard product, and $\mathbf{\Pi}$ is the PPR matrix capturing higher-order structural proximity.
The constraint loss is then defined as:
\begin{equation}
\mathcal{L}_{\mathrm{con}} = - \sum_{i,j} \mathbf{M}_{ij} \log \frac{\exp(\mathrm{sim}(\mathbf{h}_i, \mathbf{h}_j)/\tau)}{\sum_{k \neq i} \exp(\mathrm{sim}(\mathbf{h}_i, \mathbf{h}_k)/\tau)},
\end{equation}
where $\mathbf{h}_i$ is the representation of node $i$, $\mathrm{sim}(\cdot,\cdot)$ denotes cosine similarity, and $\tau$ is a temperature parameter. 

This additional loss encourages representations of structurally and semantically related nodes to be closer, while pushing apart unrelated ones. 
In this way, the model not only learns to distinguish positive and negative links correctly but also captures rich community-level and multi-hop structural patterns. 

Therefore, the overall training objective is given by:
\begin{equation}
\mathcal{L} = \mathcal{L}_{\mathrm{CE}} + \alpha \mathcal{L}_{\mathrm{con}},
\end{equation}
where $\alpha$ balances the contributions of the two components.

\subsection{Complexity Analysis}

We analyze the computational complexity of the proposed CELP framework, which consists of three main modules: the Global Enhancement Module, the Structure Enhancement Module, and the Local Representation Enhancement Module.

\paragraph{Global Enhancement Module}
We first apply a community detection algorithm to partition the graph into $K$ disjoint communities. We adopt FluidC for its scalability, which runs in near-linear time $\mathcal{O}(|E|)$ in practice. Afterward, for each community center $c_k$, we perform a BFS traversal on the entire graph to compute the shortest-path distances from $c_k$ to all nodes. 
Each BFS has time complexity $\mathcal{O}(|V| + |E|)$, resulting in a total cost of $\mathcal{O}(K(|V| + |E|))$. 
This step is fully precomputable and executed only once.

\paragraph{Structure Enhancement Module}
This module constructs a candidate edge set by filtering node pairs that belong to the same community and exhibit high global importance measured by PageRank scores. The PageRank vector can be computed iteratively in $\mathcal{O}(k|E|)$ time, where $k$ is the number of iterations. 
We then pre-train a lightweight link prediction model 
$F(u,v \mid G)$ to estimate connection probabilities.
Assuming the base GNN has per-epoch complexity $\Omega_b(V, E)$, this stage incurs $T \cdot \Omega_b(V, E)$ time for $T$ epochs. Based on the predicted scores, confidence-based filtering is applied to obtain the final edge addition set $E_{\text{add}}$ and removal set $E_{\text{rm}}$, which involves only linear scans over candidate pairs and introduces negligible overhead.

\paragraph{Local Representation Enhancement Module}
In this module, we extract neighborhood, path-based, and community-aware features for target node pairs. To avoid explicit computation of high-order adjacency powers, we employ an efficient path approximation strategy based on Personalized PageRank. 
Assuming $t$ target links, maximum node degree $d_{\max}$, hop size $r$, and feature dimension $F$, the overall complexity of this module is $\mathcal{O}(|V|d_{\max}^r + t d_{\max}^r F)$. 
By processing subgraphs in batches rather than per edge, the computational cost is significantly reduced compared to methods such as SEAL~\cite{zhang2018link}.

Overall, the computational cost of CELP is dominated by sparse graph operations and a small number of GNN forward passes. The global module is fully precomputable, the structure module introduces limited amortized overhead, and the local module 
leverages efficient approximations and batch processing, enabling CELP to scale to large graphs.

\section{Experiments}
In this section, we evaluate the effectiveness of the proposed framework CELP for link prediction on graphs, conducting extensive experiments on multiple datasets of varying scales to address the following questions:
Q1: How does CELP perform compared to state-of-the-art models?
Q2: What is the contribution of each component in the CELP framework to overall performance?
Q3: How sensitive is CELP to hyperparameters?

\subsection{Performance on Link Prediction}

\subsubsection{Datasets}

We evaluate the performance of different models on 6 widely-used public datasets, which includes (i) citation networks: Cora, CiteSeer, PubMed; (ii) co-purchase networks: Photo, Computers; (iii) OGB datasets: Collab

The citation networks (Cora, CiteSeer, and PubMed) represent academic citation graphs, where nodes correspond to research papers and edges denote citation relationships between papers. These datasets are characterized by relatively sparse connectivity and clear community structures that often align with research topics or scientific domains, making them suitable benchmarks for evaluating structural representation learning and link prediction performance.

The co-purchase networks (Photo and Computers) are derived from Amazon product data, where nodes represent products and edges indicate that two products are frequently purchased together. Compared to citation networks, these graphs are generally larger and more densely connected, with more heterogeneous and overlapping communities, posing greater challenges for link prediction models in capturing meaningful structural patterns.

The Collab dataset from the Open Graph Benchmark (OGB) is a large-scale collaboration network, in which nodes represent authors and edges indicate co-authorship relationships. This dataset exhibits complex, evolving graph structures and high-degree variability, making it a challenging benchmark for assessing the scalability and robustness of link prediction methods on real-world graphs.

\subsubsection{Baselines}

For a fair comparison, we selected representative models from the following three categories:
(1) Heuristic methods: Common Neighbors (CN), Adamic-Adar (AA), and Resource Allocation (RA);(2) Node-level graph neural networks: Graph Convolutional Network (GCN) and GraphSAGE;(3) Link-level prediction models: SEAL \cite{zhang2018link}, NeoGNN \cite{yun2021neo}, BUDDY \cite{chamberlain2022graph}, MPLP \cite{dong2024pure}, and NCNC \cite{wang2023neural}.
Each experiment is conducted 10 times, with the average score and standard deviations reported. We adopt different evaluation thresholds for small-scale and large-scale datasets to ensure fair and meaningful comparison across varying graph sizes and densities. Specifically, we use HR@100 (Hit Ratio at rank 100) on small-scale datasets such as Cora, Citeseer, and Pubmed, while employing HR@50 on large-scale datasets including Photo, Computers, and Collab.

This design choice is based on the observation that large-scale graphs typically have significantly more candidate negative links, making the top-$k$ ranking task more challenging. A smaller $k$ thus provides a stricter and more realistic evaluation. Conversely, for smaller datasets with fewer nodes and candidate edges, HR@100 ensures sufficient resolution to distinguish model performance. Using different $k$ values accordingly balances evaluation difficulty and maintains metric sensitivity across datasets of different scales.

\subsubsection{Results}
To answer Question Q1, Table~\ref{tab:link_prediction_results} presents a comprehensive comparison of CELP against a wide range of baseline models across six benchmark datasets. Overall, CELP consistently achieves the best performance on all datasets, demonstrating its superior capability in capturing both local and global graph structural information.

We first focus on the citation networks (Cora, Citeseer, and Pubmed). On these datasets, CELP significantly outperforms traditional heuristic-based methods (CN, AA, RA) as well as vanilla GNN-based models such as GCN and GraphSAGE. For example, on Cora, CELP achieves an HR@100 of 93.34, exceeding the second-best method NCNC by 3.69 points. Similar performance gains can be observed on Citeseer and Pubmed, where CELP reaches 95.41 and 84.11 respectively, outperforming all competing methods. These improvements indicate that CELP can capture higher-order structural dependencies that are difficult to model using purely local aggregation or handcrafted similarity measures.

Notably, the improvements are more pronounced on smaller-scale datasets such as Cora and Pubmed, whereas the performance gap on larger networks like Computers and Collab is relatively smaller. This can be attributed to the clearer community structures in smaller graphs: nodes tend to form semantically coherent clusters, allowing CELP’s distance encoding and community-aware aggregation mechanisms to generate more informative structural representations. In contrast, in larger networks, community boundaries are often noisier or more overlapping, which may slightly reduce the relative advantage of community-based modeling, though CELP still maintains the top performance.

On larger and more complex graphs, including Photo, Computers, and Collab, CELP also consistently achieves the best performance. For instance, on Photo, CELP attains an HR@50 of 58.87, slightly surpassing MPLP, while on Computers and Collab, CELP improves upon the strongest baselines with HR@50 scores of 43.81 and 67.37, respectively. These results suggest that CELP remains effective even when community structures become noisier or more overlapping, benefiting from its balanced design that combines local neighborhood aggregation with global community-aware structural encoding.

In summary, these results validate CELP’s design choices: the integration of local neighborhood aggregation with community-aware global structural encoding consistently improves link prediction accuracy. The model is especially effective in small, well-structured graphs, yet it also demonstrates competitive performance and stability on larger, more complex networks, making it a versatile approach for link prediction tasks.

\begin{figure}[t]
\centering
\includegraphics[width=0.75\textwidth]{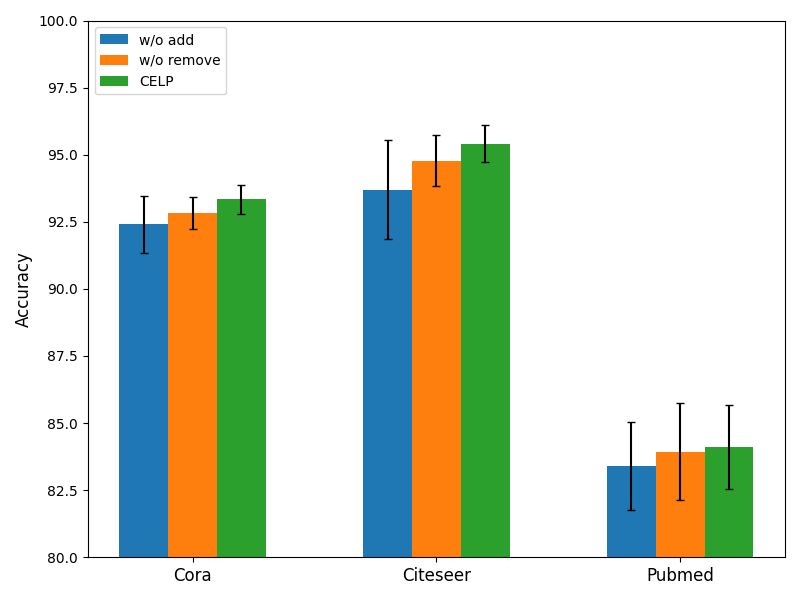}
\caption{Ablation study results of CELP on the structure enhancement module. 
``w/o add'' denotes removing the edge addition strategy, and ``w/o remove'' denotes removing the edge removal strategy.}
\label{fig:add_remove}
\end{figure}

\subsection{Ablation Studies}

To evaluate the impact of each key component on the overall performance of CELP (Q2), we conduct an ablation study by selectively removing individual modules. 
The results are shown in Table~\ref{tab:ablation_studies}. The results on the Cora, Citeseer, and Pubmed datasets use the same evaluation metric as before: HR@100.  
The results on the Photo, Computers, and Collab datasets use the evaluation metric HR@50.
''w/o GE'' refers to the removal of the Global Enhancement Module, thus excluding global structural features; 
''w/o SE'' refers to removing the Structure Enhancement Module and its edge completion mechanism; and 
''w/o LE'' means removing the Local Representation Enhancement Module and relying only on the original node and neighborhood features.

\begin{table}[t]
\centering
\renewcommand\arraystretch{1.3}
\caption{Ablation study results of CELP with different community detection strategies.}
\label{tab:community_detection}

\begin{tabular}{l c c}
\hline
Dataset & FluidC & Louvain \\
\hline
Cora      
& 93.34 $\pm$ 0.54 
& 92.74 $\pm$ 1.26 \\

Citeseer  
& 95.41 $\pm$ 0.69 
& 93.54 $\pm$ 1.42 \\

Pubmed    
& 84.11 $\pm$ 1.57 
& 82.51 $\pm$ 1.32 \\

Photo     
& 58.87 $\pm$ 2.98 
& 56.69 $\pm$ 2.63 \\

Computers 
& 43.81 $\pm$ 2.95 
& 40.84 $\pm$ 3.27 \\

Collab    
& 67.37 $\pm$ 0.56 
& -- \\
\hline
\end{tabular}
\end{table}

The results demonstrate that all three enhancement modules contribute significantly to the model’s performance. 
The Global Enhancement Module provides global structural awareness through community features, which proves especially beneficial on small-scale graphs such as Cora, Citeseer, and Pubmed, where sparse adjacency structures limit local information, making global context more valuable. 
The Structure Enhancement Module improves connectivity via edge augmentation without introducing notable noise. 
The Local Representation Enhancement Module enhances local structural modeling between node pairs via path-based features, showing greater effectiveness on large-scale graphs like Photo, Computers, and Collab, where rich and complex local topologies benefit from fine-grained path-based structural representation.

In addition to the ablations on the three main enhancement modules, we further investigate the individual contributions of the two components within the Structure Enhancement Module: the edge addition strategy and the edge removal strategy. As illustrated in Figure~\ref{fig:add_remove}, we conduct separate ablation studies by disabling each component in turn, denoted as ''w/o add'' and ''w/o remove'', respectively. 

\begin{figure}[t]
\centering
\includegraphics[width=0.75\textwidth]{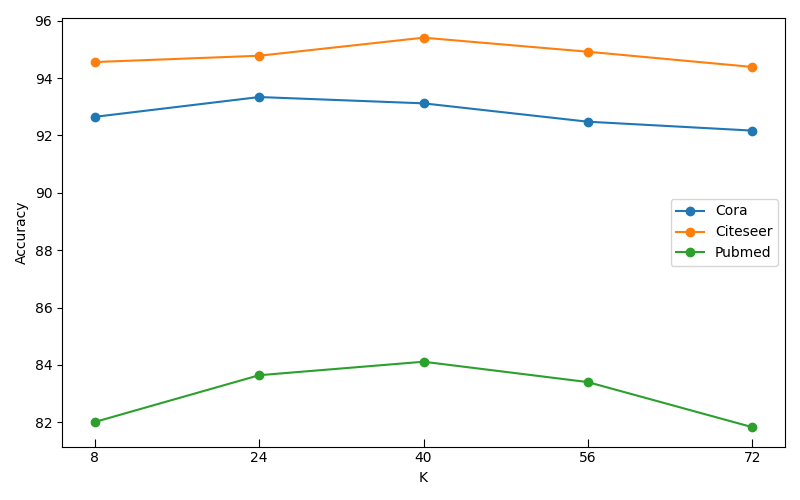}
\caption{The impact of the number of communities on model performance.}
\label{fig:num_com}
\end{figure}

The experimental results reveal that both strategies contribute positively to model performance, validating the effectiveness of structural refinement through edge adjustment. Interestingly, across all three benchmark datasets, the \textit{remove} strategy shows a more substantial impact than the \textit{add} strategy. This observation suggests that, in real-world graph data, eliminating spurious or noisy edges may be more critical than recovering missing ones. Given that many graph datasets are constructed through automated or imperfect processes, reducing noise can help sharpen structural signals and improve generalization, especially in link prediction tasks where edge quality is crucial.

Moreover, we conduct an additional ablation study by replacing the FluidC algorithm with the widely-used Louvain method for community detection. As shown in Table~\ref{tab:community_detection}, our model consistently achieves better performance when using FluidC across all datasets. One key advantage of FluidC lies in its ability to explicitly control the number of communities, enabling a more balanced trade-off between structural granularity and model complexity, which better serves the downstream link prediction task.

In contrast, the Louvain algorithm automatically determines the number of clusters based on modularity maximization, which often results in inconsistent and suboptimal community partitions across different datasets. This lack of control makes it difficult to maintain uniform enhancement effects. Furthermore, Louvain could not produce results on some large-scale graphs like \textit{Collab} due to resource limitations. Therefore, we conclude that FluidC provides a more effective and flexible solution for community-aware enhancement within our framework.

\begin{figure}[t]
\centering
\includegraphics[width=0.75\textwidth]{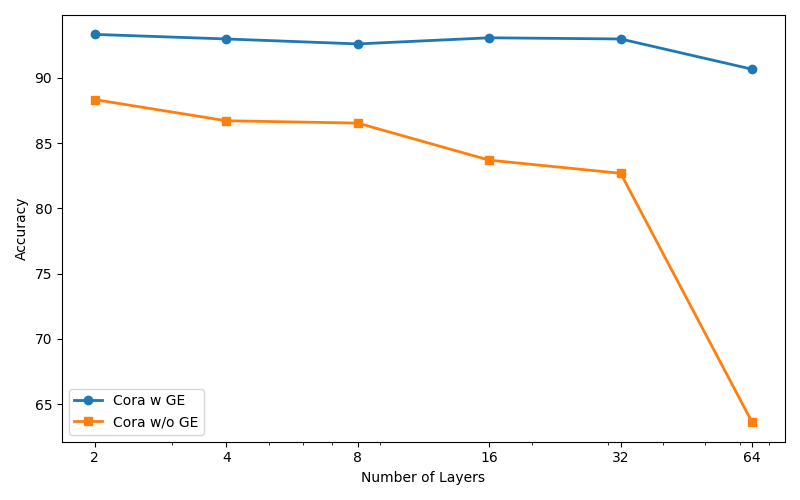}
\caption{The impact of the number of layers on model performance.}
\label{fig:layers}
\end{figure}

\subsection{Parameter Analysis}
In this section, we investigate the sensitivity of CELP to its hyperparameters (Q3). Figure~\ref{fig:num_com} illustrates how model performance varies with the number of communities $K$ used in the Global Enhancement Module. As observed, the model accuracy initially improves as $K$ increases, reaches a peak, and then declines as $K$ continues to grow. This trend is especially pronounced on the Pubmed dataset, indicating that the community granularity plays a crucial role in capturing meaningful global structure.

\begin{figure}[t]
\centering
\includegraphics[width=0.75\textwidth]{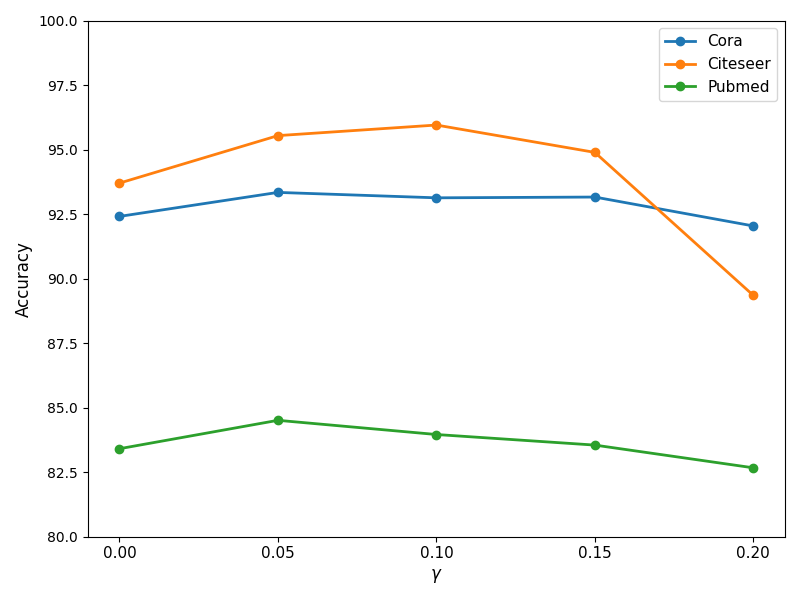}
\caption{Parameter sensitivity analysis of the edge addition ratio $\gamma$.}
\label{fig:gamma}
\end{figure}

A smaller number of communities may cause overly coarse partitioning, where diverse regions of the graph are merged together, thus weakening the discriminative power of the community features. On the other hand, an excessively large number of communities results in very small or fragmented groups, which often lack structural coherence. This not only introduces noise but also increases the risk of overfitting, as the community labels become less reliable indicators of global context.

Therefore, selecting an appropriate number of communities is critical for balancing global expressiveness and structural stability. These results emphasize the importance of tuning $K$ based on dataset characteristics, such as graph size and density, to achieve optimal performance in representation learning and link prediction tasks.

To further verify the effectiveness of the Global Enhancement Module, we conduct an experiment to analyze the model's performance under varying GNN layer depths. Specifically, we vary the number of GNN layers from 2 to 64 on the Cora dataset, and compare the results between the model with GE and the one without. As shown in Figure~\ref{fig:layers}, the model equipped with GE consistently outperforms its counterpart across all layer settings.

Notably, the performance of the model without GE deteriorates significantly as the number of layers increases, dropping from 88.34 at 2 layers to 63.60 at 64 layers. This sharp decline indicates the classical over-smoothing phenomenon, where node representations become indistinguishable due to excessive message passing. In contrast, the model with GE maintains relatively stable performance even at deeper layers, demonstrating strong resistance to over-smoothing.

These results suggest that introducing global community-level features via the GE module provides a complementary global context that preserves discriminative capacity, thereby mitigating the negative effects of deep stacking in GNNs. This confirms the benefit of integrating global structural information to improve model robustness and scalability in deeper architectures.

\begin{figure}[t]
\centering
\includegraphics[width=0.75\textwidth]{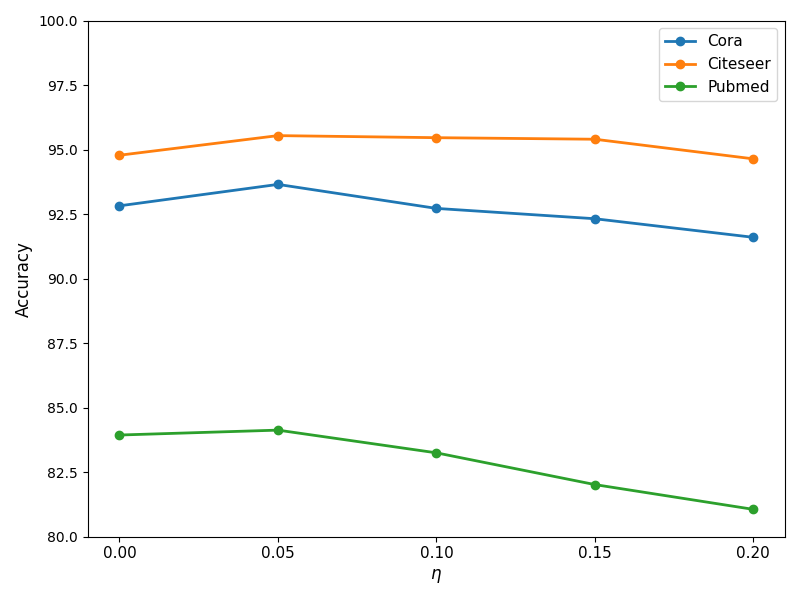}
\caption{Parameter sensitivity analysis of the edge removal ratio $\eta$.}
\label{fig:eta}
\end{figure}

We investigate the sensitivity of CELP to the edge completion ratio $\gamma$ and the edge pruning ratio $\eta$. The experimental results are shown in Figure~\ref{fig:gamma} and Figure~\ref{fig:eta}, respectively.

Figure~\ref{fig:gamma} illustrates the effect of varying $\gamma$, which controls the proportion of high-confidence edges added from the candidate set. When $\gamma = 0$, CELP degenerates to a setting without edge completion and achieves 
relatively inferior performance. As $\gamma$ increases from 0 to a small value (e.g., $\gamma = 0.05$ or $0.1$), performance consistently improves across all datasets. This trend indicates that introducing a limited number of reliable 
edges helps recover missing structural information and enhances graph connectivity, thereby facilitating more effective message passing.

When $\gamma$ becomes larger (e.g., $\gamma \geq 0.15$), performance starts to decline. This degradation suggests that excessive edge addition may introduce spurious or redundant connections, which distort the underlying graph structure 
and introduce noise into the learning process. These results highlight the need for conservative edge completion and demonstrate that a moderate $\gamma$ achieves the best balance between structural enrichment and noise control.

Figure~\ref{fig:eta} presents the sensitivity of CELP to the edge pruning ratio $\eta$, which determines the proportion of low-confidence edges removed from the original graph. With a small $\eta$, pruning a limited number of unreliable edges 
leads to consistent performance gains, indicating that removing noisy or potentially mislabeled edges improves the structural quality of the graph. This denoising effect enables the model to focus on more informative relational patterns during training.

As $\eta$ increases further, performance gradually deteriorates. Over-pruning can remove informative or semantically meaningful edges, resulting in excessive graph sparsification and the loss of critical structural signals. Consequently, the benefits of noise reduction are offset by information loss.

Taken together, the results in Figure~\ref{fig:gamma} and Figure~\ref{fig:eta} demonstrate a clear trade-off: both edge completion and edge pruning are beneficial when applied moderately, while overly aggressive structural modifications are detrimental to performance. This observation supports the design of CELP, which combines cautious edge addition with limited edge removal to improve robustness and generalization.

\begin{figure}[t]
\centering
\includegraphics[width=0.75\textwidth]{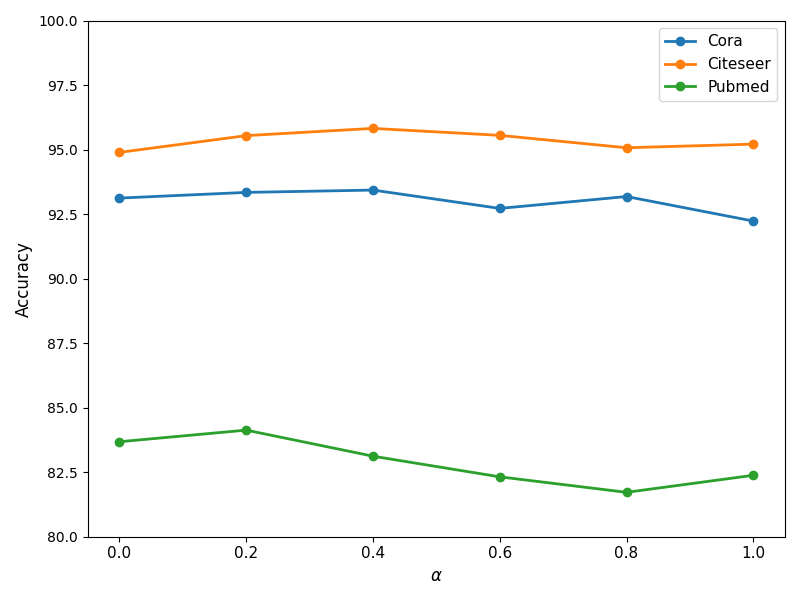}
\caption{Parameter sensitivity analysis of the loss weight $\alpha$.}
\label{fig:alpha}
\end{figure}

We further examine the effect of the weighting factor $\alpha$, which controls the contribution of the contrastive objective during training. The corresponding results are shown in Figure~\ref{fig:alpha}.

As shown in Figure~\ref{fig:alpha}, incorporating the contrastive loss with a moderate weight consistently improves performance compared to omitting it. In particular, relatively small values of $\alpha$ (e.g., $\alpha = 0.2$ or $0.4$) achieve strong and stable results across datasets, indicating that the contrastive objective provides effective auxiliary supervision for enhancing representation discrimination.

In contrast, further increasing $\alpha$ does not yield additional gains and may even degrade performance. This suggests that over-emphasizing the contrastive objective can interfere with optimizing the primary link prediction task. A likely explanation is that an excessively large contrastive weight disrupts the balance between the auxiliary and task-specific objectives, causing the contrastive loss to dominate the optimization process.

These results indicate that the contrastive objective is most effective when used as a lightweight regularizer rather than a dominant training signal. Accordingly, we adopt moderate values of $\alpha$ (e.g., $0.2$ or $0.4$) in all experiments.

\section{Conclusions}
In this work, we studied the role of community structure in improving link prediction and proposed a Community-Enhanced Link Prediction (CELP) framework that explicitly incorporates community-level information into both graph structure refinement and representation learning. CELP enhances the input graph via community-aware, confidence-guided edge completion and pruning, and jointly integrates multi-scale structural features to capture both local connectivity
patterns and global topological relationships.

Extensive experiments on multiple benchmark datasets demonstrate that CELP consistently outperforms strong baselines, validating the effectiveness of community-aware structural enhancement for link prediction. These results
highlight that leveraging community structure as a global inductive bias can significantly improve the robustness and accuracy of link prediction models. We hope this work will encourage further exploration of community-driven structural modeling in graph representation learning.

\bibliographystyle{elsarticle-num}
\bibliography{references}
\end{document}